\documentclass[12pt,preprint]{aastex}

\def\hide#1{}

\begin{document}
\title{Hard X-ray Flares Preceding Soft X-ray Outbursts in Aquila X-1: A Link between Neutron Star 
and Black Hole State Transitions }
\author{Wenfei Yu, Marc Klein-Wolt, Rob Fender and Michiel van der Klis}
\affil{Astronomical Institute, ``Anton Pannekoek'', University of Amsterdam, 
Kruislaan 403, 1098 SJ Amsterdam, The Netherlands. E-mail: yuwf,klein,rpf,michiel@astro.uva.nl}

\begin{abstract}
We have analyzed {\it Rossi X-ray Timing Explorer} (RXTE) data of 
the neutron star transient Aquila X-1 obtained 
during its outbursts in May/June 1999 and 
September/October 2000. We find that in the early rise 
of these outbursts, a hard flare in the energy range above 15 keV 
preceded the soft X-ray peak. The hard X-ray flux of the hard flares at maximum 
was more than a factor of three stronger than at any other point 
in the outbursts. The rise of the hard X-ray flare to this maximum, 
was consistent with a monotonically 
brightening low/hard state spectrum. After the peak of the hard flare, a sharp 
spectral transition occurred with spectral pivoting in the range 8--12 keV. 
Our timing analysis shows that during the hard flare the power spectra were 
mainly composed of band-limited noise and a $\sim$ 1--20 Hz QPO, which 
correlate in frequency. Immediately after the hard flare, the power spectra 
turned into power law noise. The spectral and timing properties during 
and after the hard flares are very similar to those in black hole transients 
during the early rise of an outburst. We suggest that these hard flares and 
spectral transitions in Aql X-1 are of the same 
origin as those observed in black hole transients. This leads to the 
association of the 1--20 Hz QPOs and band-limited noise in Aql X-1 with 
those in black hole transients. We discuss the impact of this discovery on 
our understanding of soft X-ray transient outbursts, state 
transitions and variability in X-ray binaries.
\end{abstract}


\section{Introduction}
Soft X-ray Transients (SXTs) undergo outbursts in 
which their luminosity increases by several orders of magnitude from quiescent
luminosity. From the quiescence to the outburst peak, black hole (BH) SXTs may go 
through a low/hard state in the rising phase of their outburst to an intermediate 
state, a high/soft state, or even a very high state. A hard flare in the rising phase of the 
outburst has been observed in the 1998 outbursts of XTE J1550-564 (Wilson \& Done 2001; 
Hannikainen et al. 2001) and 4U 1630-47 (Hjellming et al. 1999), and the 1999 
outburst of XTE J1859+226 (Brocksopp et al. 2002; Hynes et al. 2002). Brocksopp 
et al. (2002) suggest that, if not all, many BH SXT outbursts may 
be preceded by such a hard flare. 
The corresponding timing feature is a quasi-periodic oscillation (QPO) in 
the frequency range from about 0.1 Hz to above 10 Hz whose frequency increases 
during the rise, in many cases accompanied by harmonics (e.g., XTE J1550-564: 
Cui et al. 1999; 4U 1630-47: Dieters et al. 2000; XTE J1859+226: Markwardt, 
Marshall \& Swank 1999). It is not yet known if both preceding hard flares and QPOs 
occur in the outburst rise of neutron star SXTs. 
 
Aquila X-1 is one of the best-known 
neutron star low-mass X-ray binary (LMXB) SXTs. It 
shows X-ray spectral and timing properties similar to those of persistent 
LMXBs and was identified as an `atoll' source (Reig et al. 2000). However, 
the spectral variations are complicated at lower luminosities 
(Muno, Remillard \& Chakrabarty 2002 ; Gierlinski \& Done 2002; 
Maccarone \& Coppi 2002, here after MRC02, GD02, and MC02, respectively), 
and evidence for a `propeller' state was found 
(Zhang, Yu \& Zhang 1998; Campana et al. 1998). Spectral states 
in neutron star SXTs, for instance Aql X-1, may be composed of 
persistent neutron star LMXB states (e.g. characterized in the 
color-color diagram) and `propeller' states.     
 
In this Letter, we report the 
discovery of preceding hard flares in the 1999 and the 2000 outbursts 
with the observations of High Energy X-ray Timing Experiment (HEXTE) on board 
the RXTE. This is the 
second time a preceding hard flare is seen in an outburst of a neutron star 
soft X-ray transient (Cen X-4: Bouchacourt et al. 1984). In addition, we also find that the associated 
timing/spectral states and the state transition in Aql X-1 are very similar 
to those of black hole transients during the outburst rise.

\section{Observations and Results}
The early rises of the two outbursts were fully covered with RXTE pointed observations. 
In the observations of the 1999 outburst (Observations 40047 and 40049), the PCA data 
used for the timing analysis were from the {\it GoodXenon1\_2s} and {\it GoodXenon2\_2s} modes. 
In those of the 2000 outburst (Observation 50049), {\it E\_125us\_64M\_0\_1s} 
was used. We also used the light curves and spectra from the standard products 
generated by FTOOLS v.5.1. 

\subsection{RXTE Daily Average Light Curves}
First, we calculated daily average light curves from definitive 1 dwell 
ASM light curve and the PCA (2--9 keV, 9--20 keV and 20--40 keV) and HEXTE (15--30 keV 
and 30--60 keV) standard products. We found that 
the PCA (20--40 keV) and HEXTE (15--30 and 30--60 keV) light curves reached 
their peaks before the ASM peak rates, with peak rates around 3.5 c/s/PCU, 39.2 c/s and 
15.0 c/s for the 1999 outburst, and 5.5 c/s/PCU, 76.6 c/s and 23.8 c/s 
for the 2000 outburst, respectively. The ASM and the HEXTE (15--30 keV) cluster A light curves 
are shown in Fig.1. For both outbursts, the 
HEXTE flux reached its peak when the ASM count rate was about 1/3 of 
the ASM peak rate, then decreased and remained at about 1/3 of its peak flux during the 
outburst peak, and finally decreased significantly in the outburst decay. 
\subsection{Spectral Analysis}
The PCA energy spectra of the first 9 (or 8) pointings (see Fig.~1), corresponding to the 
outburst rise and state transition were studied. We used the standard spectra 
for observations with no X-ray bursts. Whenever there was a burst, we 
generated energy spectra, excluding the bursts, 
from {\it Standard 2} data with Ftools v5.2. We applied the new Combined Models (CM) 
for the background. The nine energy spectra for the 1999 outburst are shown in Fig.~2. 
As seen in the first eight spectra in time sequence, the 10--60 keV X-ray flux grew 
systematically. When the ASM rate reached its peak two days later, the spectrum 
became much softer (9th spectrum). We identify the top two energy spectra, corresponding 
to the 8th and the 9th pointings, as indicative of two distinct spectral states 
with a spectral transition between them. Our detailed analysis using XSPEC v11.2.0 shows 
that these spectra are very similar to the hard and 
soft state spectra in black hole transients. The energy spectra 
between 3 keV and 60 keV can be described with the following models, all 
with a reduced $\chi^{2}$ $\le$ 2.5: 1) a power law (PL) component 
(Pointings 1-2 of the 1999 outburst (or pointings 1-2 of the 2000 outburst)) 
2) a PL plus a blackbody (BB)(Pointings 3-4 (3)), 
3) A PL+BB plus a Gaussian line around 6.5 keV (Pointings 5-7 (4)), 
and 4) a cutoff-PL plus a BB and a Gaussian line around 6.5 keV (Pointings 8-9 (5-8)). 
All models include an absorption of $N_{H}\sim5.0\times{10}^{21} atom~{cm}^{-2}$. 
The power law indices corresponding to 
the hard flare in the 1999 outburst were in the range 1.6--1.8 with a typical 
1$\sigma$ error of 0.05. The power law indices of the 2000 outburst are in the 
range between 1.8 and 2.0. Both flares show the pivoting 
energy of the last two spectra to be at about 8--12 keV. In both outbursts, the 
energy spectra after the transition, if characterized by a single PL, have a photon 
index of about 3.0.

\subsection{Timing Analysis}
We calculated average Fourier power spectra corresponding to 
each RXTE pointing, excluding any X-ray burst. The 
Fourier transforms were performed on each 256 s segment of the event mode data from PCA, 
which were re-binned to achieve a Nyquist frequency of 1024 Hz. In Fig.~3, 
we show the power spectra during the hard flare 
peak (`hard') and immediately after the hard flare (`soft') for the 
1999 outburst, corresponding to the 7th and the 9th pointings. 
The power spectra corresponding to 
the hard flare can be described as a band-limited noise (BLN) plus a QPO feature on the slope. 
In general, we can use 2 or 3 Lorentzians to fit the average power spectra with a 
reduced $\chi^{2}$ around 1.0. We only focus on 
the two Lorentzians which represent the two dominant components, namely BLN and QPOs. 
The fractional rms of the QPO was in the range between 11\% and 17\%. The BLN fractional rms 
was 12--15\%. The power spectra immediately after the hard flare 
are well fit by power law noise, with a fractional rms 
less than 4\%. This trend of strong decrease of the BLN and increase of the power-law noise 
is consistent with that in the transition from `island' state to `banana' state. 
As noted previously (van der Klis 1994), the transition of the power spectra shows remarkable 
similarities to those observed in the transition from low to high states of 
black hole X-ray binaries (see e.g. Mendez, Belloni \& van der Klis 1998). 
In Fig.~4, we plot the correlation between the break frequency ${\nu}_{bk}$ and the 
QPO frequency ${\nu}_{qpo}$ for the 1999 outburst of Aql X-1. The correlation is in 
good agreement with the main correlation track for other black hole and neutron 
star sources (Wijnands \& van der Klis 1999; Psaltis, Belloni \& van der Klis 1999; 
Belloni, Psaltis, \& van der Klis 2002, here after BPK02). 

Using HEXTE light curve above 15 keV, which is higher than the 
pivoting energy, is crucial for the detection of preceding hard flares 
compared with previous X-ray color analysis of the same observations (MRC02; GD02; MC02). 
Furthermore, we have combined both timing and spectral properties which are necessary to 
describe a source state. It is worth noting that no preceding hard flares could be seen 
in the RXTE observations before 1999 probably because of lack of coverage of the 
early rise (e.g. Cui et al. 1998).  

\section{Comparison with Black Hole SXTs}
The preceding hard flare and the timing and spectral transition in Aql X-1 
resemble the low-to-high state transition in black hole SXTs. 
In Table 1, we compare Aql X-1 with 4U 1630-47, XTE J1550-564 and XTE J1859+226. 

The ASM light curves of the black hole transients all showed the outburst peak 
sitting as an additional $\sim$ 5 day burst upon the fast rise and exponential 
decay (FRED) outburst (with a plateau in between the rise and the decay). These 
additional bursts were identified as the very high state (VHS) (4U 1630-47: Dieters et al. 2000; 
Trudolyubov, Borozdin \& Priedhorsky 2001; XTE J1550-564, Sobczak et al. 2000), which 
has not been observed in neutron star SXT outbursts. In Table 1, 
we also show the ASM plateau rate, neglecting these VHS bursts. The rise time of 
the preceding hard flare in Aql X-1 is significantly longer than those in 
XTE J1550-564 and XTE J1859+226. Both types of source reached their hard X-ray 
peaks when their ASM rates were about 1/3 of the plateau rates. It is worth noting 
that the state transition corresponding to the decay of the hard flare seems quicker 
in Aql X-1 than in black hole systems. The difference in rise and decay time of the 
hard flares may come from the inverse Comptonization of the additional soft photons 
from neutron surface. We found that the QPO and BLN
observed in Aql X-1 have similar frequency and fractional rms amplitudes 
as those in the black hole transients.

During the onset of the outbursts in the black hole transients, the soft BB 
component and the hard PL component both strengthened, while power law steepened 
(see Behavior A through B in Dieters et al. 2000; also 
Trudolyubov, Borozdin \& Priedhorsky 2001; Sobczak et al. 2000; 
Markwardt, Marshall \& Swank 1999). This is also true for the state transition in 
Aql X-1. However, the photon index of Aql X-1 remained rather stable around 1.6-2.0 
during the rise of the hard flare and changed to about 3.0 immediately after the hard flare. 
 
\section{Discussion and Conclusion}
We have found both `preceding hard flares' and 
1--20 Hz QPO in the outburst rise of a neutron star SXT. The QPO frequency 
increased during the outburst rises in correlation with the BLN break frequency. 
The correlation agrees with those in black hole transients, and falls on 
the main correlation previously suggested for both neutron star and black hole 
X-ray binaries. The timing/spectral states corresponding to the hard flare and the 
soft X-ray maximum are very similar to the low state 
and the high state in black hole X-ray binaries (see Tanaka \& Lewin 1995 
and van der Klis 1995). 
The preceding hard flares in Aql X-1 correspond to the extended `island' branch 
with mostly soft color ($\le$ 6.5 keV) variations (GD02; MRC02). 
As demonstrated in Aql X-1 (MRC02) and 4U 1705-44 (Barret \& Olive 2002; Olive, Barret 
\& Gierlinski 2003), the extended `island' branch corresponds to the `loop' or hysteresis 
effect in of the state transitions between the `island' state and the `banana' state. 
Our results are consistent with the association of `atoll' source `island' state and 
`banana' state with black hole low state and a higher state, respectively (van der Klis 1994). 

The hard flares were found to occur at a similar soft X-ray flux level relative 
to the outburst maxima in neutron star and black hole SXTs. The decay of the hard 
flare in Aql X-1 corresponded to a spectral transition from a hard state to a soft 
state, showing spectral pivoting at an energy around 10 keV. This is similar to the 
spectral transition in the outburst rise of black hole transients in which a preceding 
hard flare is found (Cui et al. 1999; Markwardt, Marshall \& Swank 1999; 
Dieters et al. 2000; Sobczak et al. 2000). During the outburst rise, in addition to 
similar energy spectra, similar power spectra are observed in neutron star and black 
hole transients. The power spectra corresponding to the hard flare in Aql X-1 show a 
similar shape as that of black hole transients in the outburst rise, i.e. dominated 
by a QPO below 20 Hz and a BLN. The QPO frequency and the BLN break correlate as those 
found in black hole transients. The fractional rms amplitudes of the QPO and the BLN 
are also consistent with those in the black hole systems. The hysteresis effect of the 
state transition in both neutron star and black hole SXTs suggests a common origin of 
the state transitions (MC02). In other words, whereas the similarities in neutron star 
and black hole timing/spectral states have been noted many time before, 
our observations of the preceding hard flare and the associated state transition 
in SXTs have established an additional link between neutron star and black hole states.  

The similar spectral/timing states and state transition 
suggest that the QPO in Aql X-1 is that observed in the outburst rise of 
black hole transients. So does it for the BLN. 
Both the QPO and the BLN in Aql X-1 are thus not generated from 
the central neutron star. The correlation between the QPO frequency and the noise 
break in Aql X-1 as shown in Fig.~4 agrees with the main correlation previously found 
in other neutron star and black hole X-ray binaries (WK99, PBK99 and BPK02). 

In Aql X-1, as well as in Cen X-4 (Bouchacourt et al. 1984) and those black hole transients 
in which a preceding hard flare was seen during an outburst rise, the hard 
flare had a flux much higher than that in the outburst decay. The hard flux is 
thought to be generated by upscattering of soft seed photons by hot electrons 
through inverse Comptonization. This asymmetry 
then reflects an asymmetry in corona luminosity evolution (corona cooling vs. 
heating) during the outburst, and is probably the cause of the hysteresis effect 
of state transition in SXTs (e.g. Miyamoto et al. 1995). 
This is supported by the recent spectral study of the 1999 Aql X-1 
outburst (Maccarone \& Coppi 2003) and the state transitions in 4U 1705-44 
(Barret \& Olive 2002). It is worth noting that the same asymmetry is seen in 
radio flaring events associated with relativistic ejections, which are much 
brighter in the initial phase of outbursts (Hjellming \& Han 1995; Fender \& Kuulkers 2001), 
suggesting a connection between the evolution of jet and corona.   
\acknowledgments
We would like to thank Rudy Wijnands for mentioning an early hard X-ray observation of Cen X-4. 
This work was supported in part by the Netherlands Organization for
Scientific Research (NWO) under grant 614.051.002. WY would like to 
thank IHEP and 973 Projects for his past research and NSFC grant 
19903004 (2000-2002). This work has made use of data obtained through 
the High Energy Astrophysics Science Archive Research Center Online 
Service, provided by the NASA/Goddard Space Flight Center.

\clearpage

\begin{figure*}
\epsscale{0.8}
\plotone{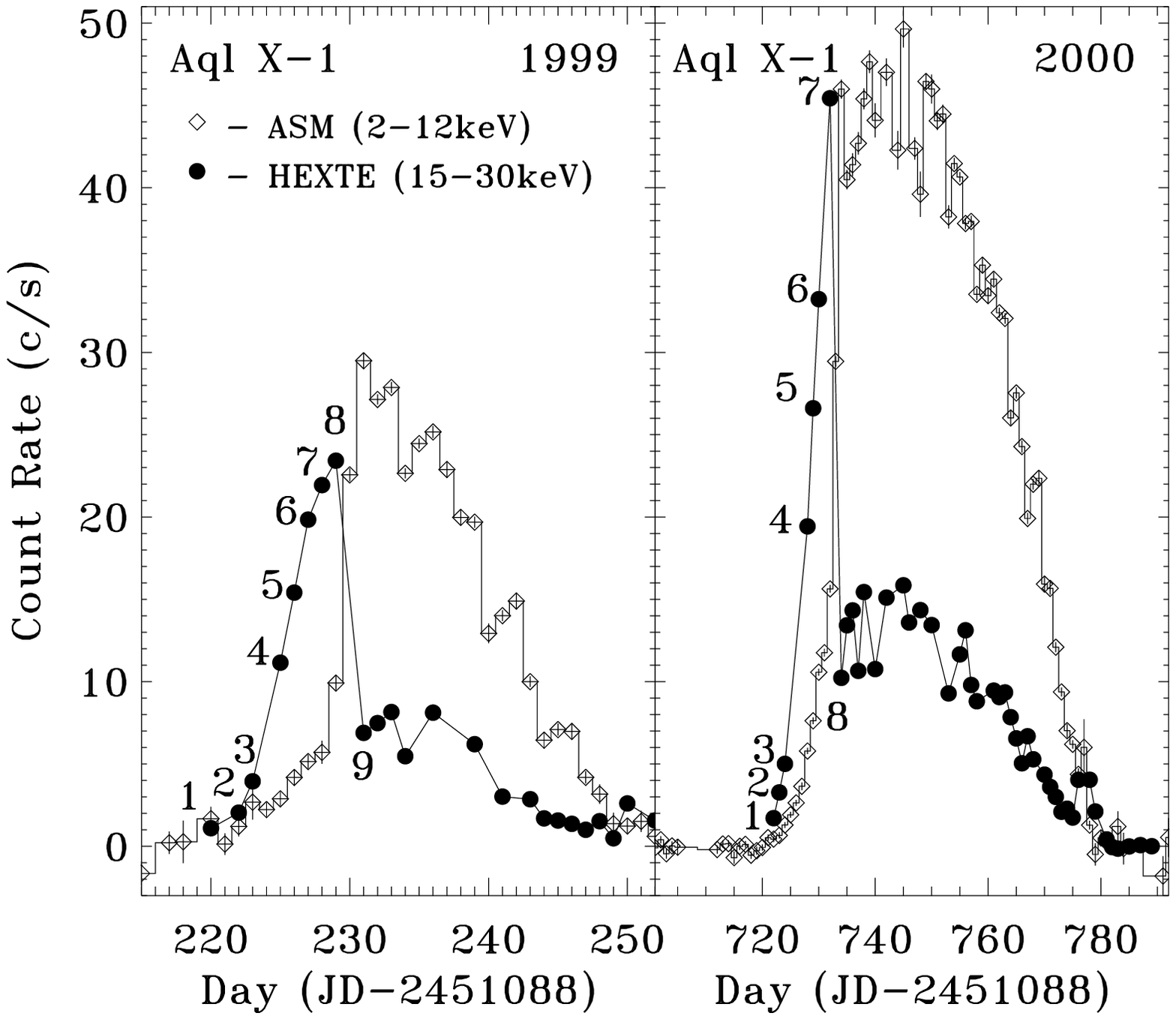}
\caption{The May/June 1999 outburst (left) and the September/October 2000 outburst (right) of 
Aql X-1. ASM light curves (2--12 keV) and the HEXTE cluster A light curves (15--30 keV) are shown 
as diamonds and filled circles, respectively. The numbers marked on the HEXTE rates 
represent those RXTE pointing observations in our analysis. }
\end{figure*}

\begin{figure*}
\epsscale{0.8}
\plotone{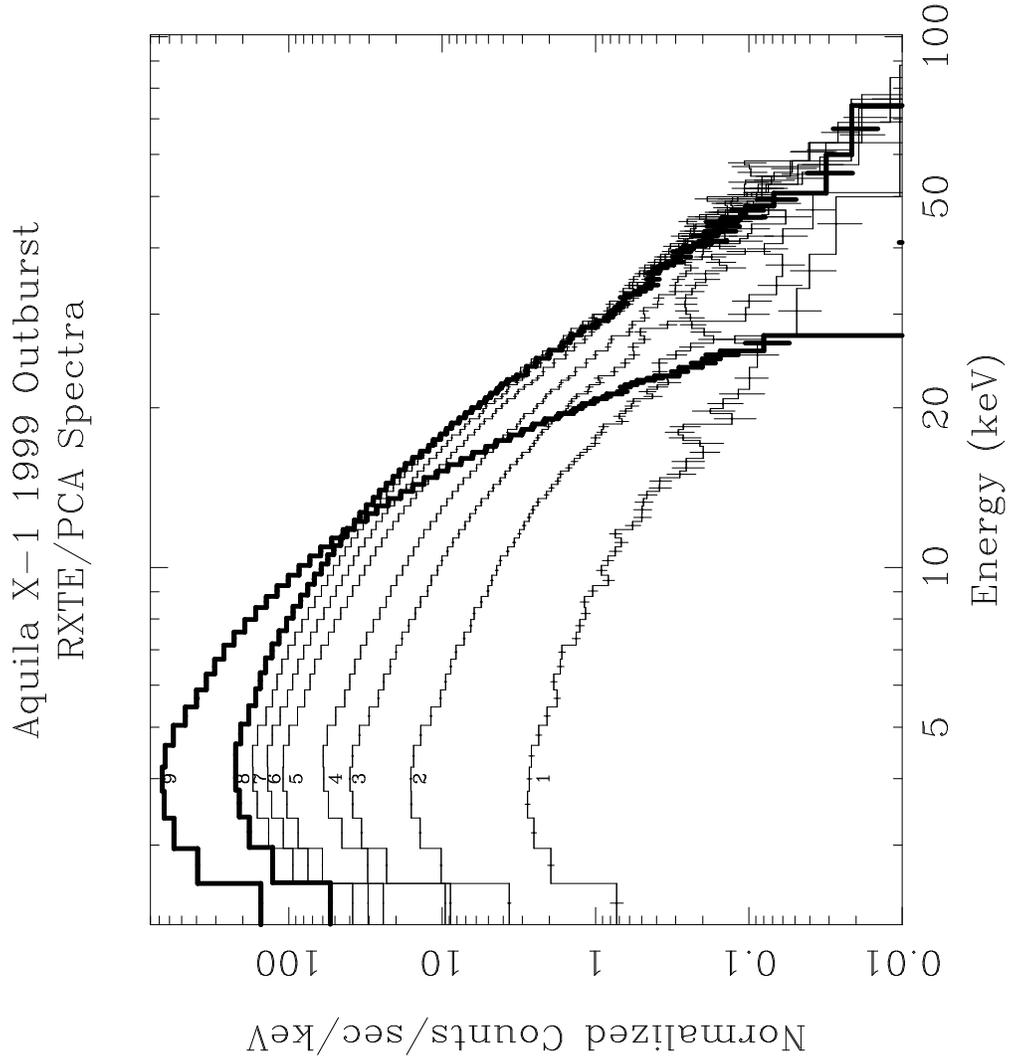}
\caption{Evolution of the RXTE/PCA energy spectra during the 1999 outburst rise. 
The first 8 spectra show an increase of the X-ray flux in the wide PCA energy band, 
corresponding to the growth of a hard flare. The ninth spectrum corresponds to 
the pointing after the transition. The pivoting energy of the top two spectra (thick)
is around 12 keV. }
\end{figure*}

\begin{figure*}
\epsscale{1}
\plotone{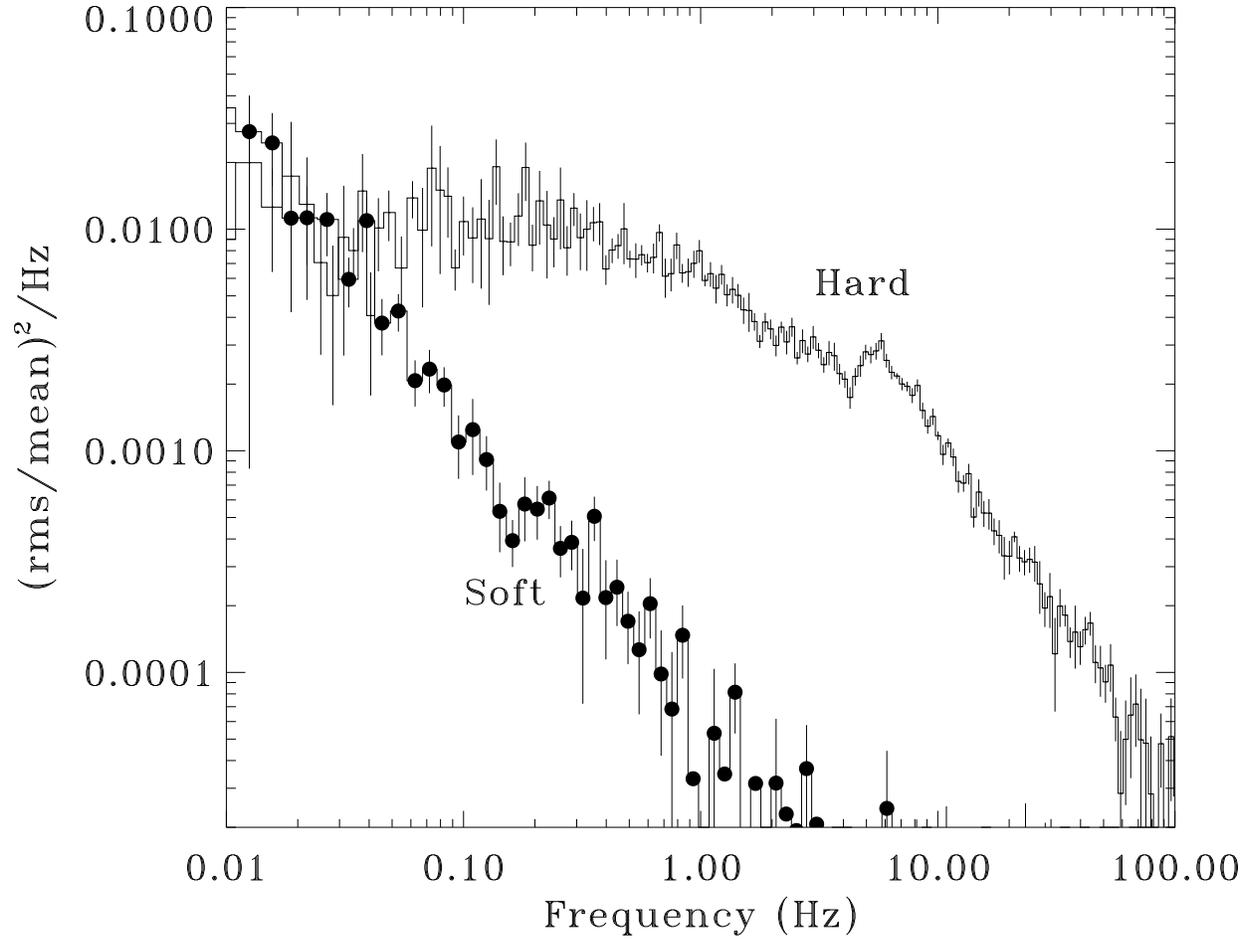}
\caption{Power spectrum showing a band-limited noise with QPOs around 5 Hz during 
the hard flare of the 1999 outburst before the transition (hard) and a power law noise 
spectrum after the transition (soft). Notice these power spectra are identical to 
those power spectra corresponding to low and high states in 
black hole X-ray binaries.}
\end{figure*}

\begin{figure*}
\epsscale{1}
\plotone{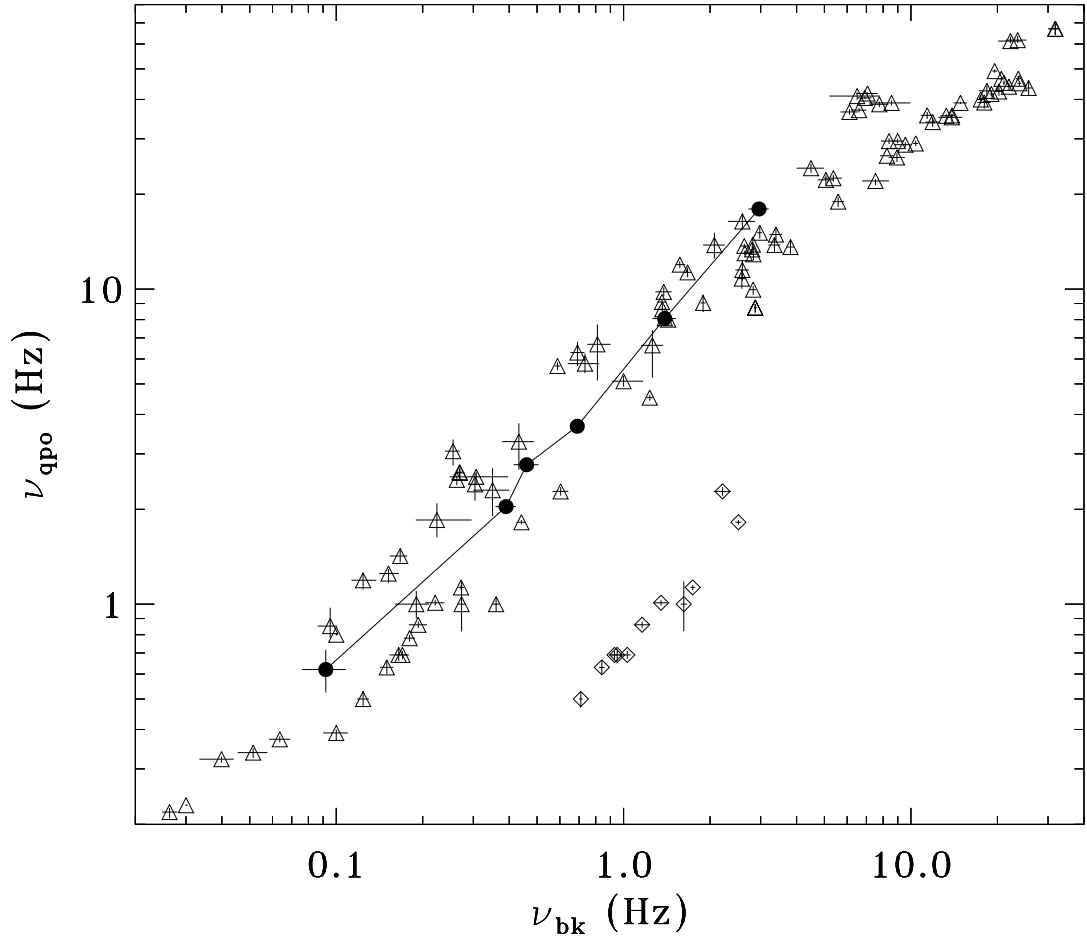}
\caption{The correlation between the break frequency and the QPO frequency for pointings 3--8 
(from lower left to upper right) during the 1999 outburst rise (marked as filled circles). 
The data overlap on the main correlation (triangles) but not on the secondary correlation 
(diamonds) for other black hole and neutron star X-ray binaries 
(from Belloni, Psaltis \& van der Klis 2002).}
\end{figure*}

\clearpage
   
\begin{deluxetable}{lrrrrrrr}
\tabletypesize{\scriptsize}
\tablecaption{Hard Flares (HFs) in the Outburst of Aql X-1 and Black Hole SXTs \label{tbl-1}}
\tablewidth{0pt}
\tablehead{
\colhead{Source} & 
\colhead{ASM Rate$^{a}$ (c/s)}   & 
\colhead{Peak Rate$^{b}$ (c/s)}  &
\colhead{Rise (Day)} &
\colhead{Decay$^{c}$ (Day)} & 
\colhead{$\nu_{QPO}$ (Hz)} &
\colhead{QPO rms} &
\colhead{PL Index}  
}
\startdata
Aql X-1/1999 &25&	10& 	9&$\le$2& $\sim$1--18&11--17\% & 1.6-2.0 \\
Aql X-1/2000 &45&	16& 10&$\le$2 & $\sim$1--6&12--15\% & 1.8-2.0 \\
4U 1630-47$^{d}$ &21& 6--16 & $\sim$14& $\le$4 & 2.7--13.7&9.2--30.7\% & 2.0--2.5 \\
XTE J1550-564$^{e}$ &120 & 40  & 2&5  & 0.08--10&17--23\% & 1.5--2.4  \\
XTE J1859+226$^{f}$ &50 & 15  & $\le$2& $\sim$3  & 0.45--10&17\% & 1.7--2.5 \\ 	 		 
\enddata

Note: (a){daily-average plateau rate (excluding probably VHS bursts)}; 
(b){The ASM rate at the HF peak}. 
(c){Decay time from the HF peak to hard X-ray plateau}. 
(d){Hjellming et al. 1999; Dieters et al. 2000; Trudolyubov, Borozdin \& Priedhorsky 2001}. 
(e){Cui et al. 1999; Hannikainen et al. 2001}.
(f){Markwardt, Marshall \& Swank 1999}
\end{deluxetable}

\end{document}